# Topological representation of layered hybrid lead halides for machine-learning using universal clusters

*Ekaterina I. Marchenko [a,b]\*, Maria G. Khrenova [c], Korolev V.V.[d], Eugene A. Goodilin [a,c], and Alexey B. Tarasov [a,c]*

[a] Faculty of Materials Science, Lomonosov Moscow State University, 119991 Moscow, Russian Federation.

[b] Department of geology, Lomonosov Moscow State University, 119991 Moscow, Russian Federation.

[c] Department of Chemistry, Lomonosov Moscow State University, 119991 Moscow, Russian Federation.

[d] MSU Institute for Artificial Intelligence, Lomonosov Moscow State University, 119192, Moscow, Russian Federation.

\* E-mail: marchenko-ekaterina@bk.ru

**Layered hybrid halide compounds offer promising functional properties, particularly tunable band gaps, conductivity, light harvesting thus making them prospective for applications in photovoltaics and optoelectronics. This study exemplifies an approach of predicting band gaps using machine learning models enhanced by invariant topological representations of these materials using the atom-specific persistent homology method in order to facilitate the discovery and design of new hybrid halide materials with tailored electronic properties.**

The design and discovery of novel materials with tailored electronic properties are crucial in advancing fields such as photovoltaics, optoelectronics, and energy storage. One such class of materials, layered hybrid lead halide compounds with perovskite-derived crystal structures (LHP), has gained significant attention due to their tunable band gaps, which are essential for optimizing their performance in electronic and energy-related applications [1–5]. However, an accurate theoretical prediction of band gaps in these materials remains a challenging task due to the complex interplay of their atomic and electronic structures [6,7].

In recent years, machine learning (ML) has emerged as a powerful tool to accelerate materials discovery by predicting key properties such as band gaps for LHP based on structural data [8–10]. It is known that the band gap depends on a number of geometric descriptors in LHP such as the metal-halogen bond length, the angles between atoms in the inorganic substructure, the layer shift factor, and some others [6,11,12]. However, modern machine learning algorithms are plausible to input information about the crystal structure written universally, for example, as a multidimensional vector. Topological representation methods, which capture the spatial arrangement and connectivity of atoms within a material, have shown a promise in enhancing ML model accuracy. In this context, the development of effective topological descriptors for hybrid halide compounds could significantly improve our ability to predict their electronic behavior. This article explores the topological representation [13] of layered hybrid lead halide compounds and its application to machine-learning models for band gap prediction.

In this study, we utilized a dataset comprising 140 two-dimensional perovskite-related crystal structures exhibiting the (100) structural type, characterized by a perovskite block thickness of n=1. This dataset[i] was sourced from [8], wherein the band gap values had previously been calculated with high precision using density functional theory (DFT) methods. Each material in the database is encoded in the Crystallographic Information File (CIF) format and is available for unrestricted access. The website provides comprehensive descriptions of material properties, including DFT-derived and experimental band gaps, chemical formulas, space groups, and other relevant data.

Figure 1 illustrates the barcodes construction for two-dimensional perovskite materials. Our methodology encompasses several essential steps. Initially, we systematically extracted various types of atoms with different crystallographic sites and their combinations within the unit cell from the crystallographic information files (.cif) contained in the dataset. Around each atom in unit cell, a sphere within a cutoff radius is constructed, into which a cloud of points (atoms) falls. Then for each sphere the number and distance between atoms are calculated, which are reflected on the barcode as lines [ii]. The number of lines means the number of bonds, and their length is the interatomic distance. Thus, we developed an atom-specific topological fingerprint representation[13] for LHP to extract detailed crystal information pertinent to machine learning applications. The topology of these structures was encoded in the form of specific barcodes. Utilizing this topological representation, we employed a gradient boosting tree (GBT)[14] model to predict the band gaps of the materials.

---

[i] The complete dataset of two-dimensional perovskites employed in this investigation is accessible via the official NMSE database website: http://www.pdb.nmse-lab.ru/.

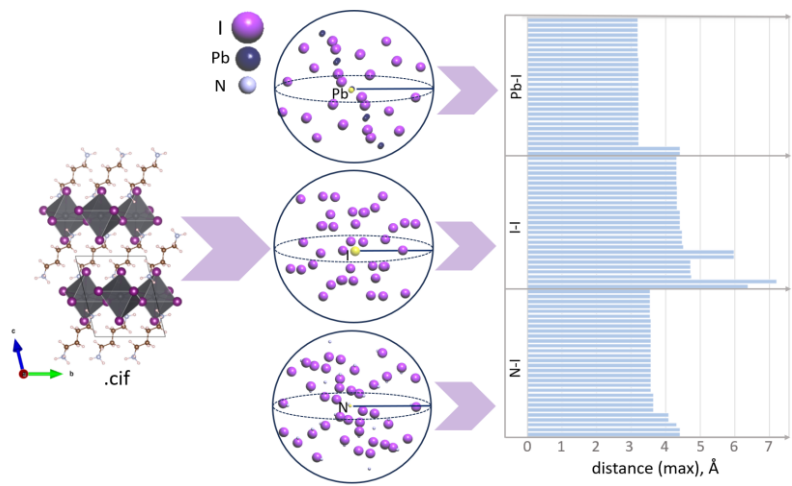

**Figure 1.** Illustration of barcodes construction of LHP crystal structure. The crystal structures were visualized using the Vesta[15] and TOPOSpro[16] programs.

The methodology and algorithm for constructing specific barcodes in Python were developed as outlined in reference [13]. The fundamental concept underlying this approach is that

---

[ii] To create a region of point clouds surrounding each atom, we employed a cutoff radius of 10 Å to generate the barcodes. For LHP structures such a radius is optimal, since interactions between adjacent layers of lead-halogen octahedra are possible up to such distances. By calculating the bond lengths from the central atom to its neighboring atoms, the resultant barcode encapsulates the geometrical and topological characteristics of the structure. In the figure 1, solid lines indicate contacts between pairs of atoms within a defined sphere; the number of lines corresponds to the frequency of these contacts in the unit cell, while the lengths of the lines represent the distances between atoms. The order of the lines in the barcode does can be chosen arbitrarily. In addition to topological information, composition-based features were incorporated, which include stoichiometric attributes reflecting elemental fractions, elemental-property statistics derived from all atoms in the crystal, and electronic structure attributes.

within a unit cell, a limited number of atoms exist, each possessing a distinct structural environment that defines its unique topological fingerprints. This approach is universal and invariant, since crystal structures descriptors are not represented as heterogeneous parameters, such as the unit cell parameters, angles and the coordinates of the atoms. All geometric parameters in this representation are presented homogeneously in the form of values of interatomic distances in the local environment of each atom. As an example, for the (100) layered hybrid lead halide crystal structure with n=1, three significant atomic pair combinations Pb-X, X-X, and N-X (where X is a halogen ion) are identified as having a substantial impact on the band structure of the material [12,17]. Changes in these three pairs of interatomic distances make a significant contribution to the change in the LHP band gaps compared to changes in the distances between other pairs of ions.

Thus, as a result of converting a classic Crystallographic Information File (CIF) of a crystal structure into a barcode, we obtain a data set that is easy to represent in a machine-readable form as a multidimensional vector due to the homogeneity of the data representation. Such information is easy to process using modern libraries for machine learning. Compared with other machine-readable crystal structure representations based on structure graphs and Coulomb matrices [18], topological descriptors using persistent homology offer the advantage of uniquely encoding structures at both local and global levels, without requiring assumptions about the underlying physics. For the selection of the machine learning (ML) algorithm, we opted for gradient boosted regression trees (GBRT) [14], to evaluate the accuracy, robustness, and efficiency of topological-based features [iii]. The performance metrics for the model predicting band gaps using topological feature vectors yielded $R^2 = 0.8$, RMSE = 0.17 eV, and MAE = 0.12 eV (refer to Figure 3). These results are consistent with contemporary machine learning models aimed at predicting the band

---

[iii] GBRT effectively integrates multiple weak predictors to formulate a robust model. The training process involves sequentially adding trees to diminish the loss function of the current model. To mitigate overfitting, each model update utilizes various randomly selected subsets of both training data and features. Hyper-parameter optimization was performed through cross-validation, evaluated using the $R^2$ metric. The hyper-parameters employed in GBRT include: *n_estimators* = 300,000, *learning_rate* = 0.001, *max_depth* = 7, *min_samples_split* = 5, *subsample* = 0.85, and *max_features* = *sqrt*. The ML models were constructed using scikit-learn software (version 0.19.2) as indicated in [19]. A tenfold cross-validation approach was utilized to validate the proposed methodology, with random data splitting repeated 20 times to assess model robustness. The median performance metrics and standard deviation across these repeated experiments were documented. Voronoi tessellations and Coulomb matrices were replicated using Magpie, which is freely available under an open-source license [20].

gaps of hybrid perovskites [8,10]. Furthermore, a commendable MAE was achieved despite the limited size of the dataset. Thus, the representation of crystal LHP structures as barcodes is a good general-purpose machine-readable representation for the targeted design of this class materials.

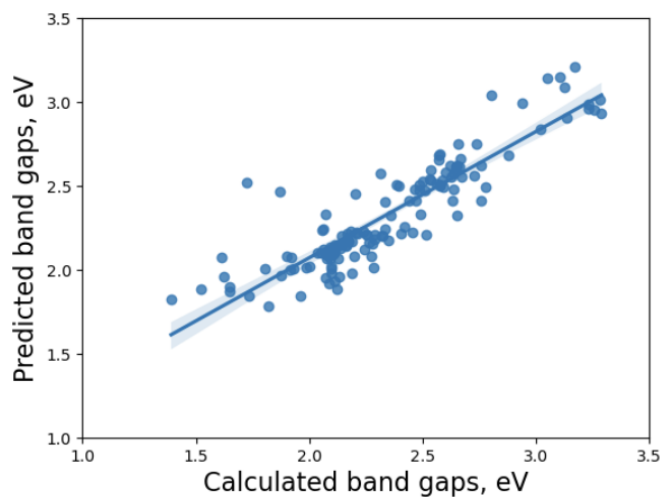

**Figure 3.** Comparison of DFT-calculated band gaps and predicted band gaps by ML algorithm for 2D hybrid lead halide materials.

Beyond LHP materials, this approach presents opportunities for addressing both the direct problem (predicting the physical properties of materials from their crystal structure) and the inverse problem (predicting crystal structures with desired properties) for other hybrid materials related to the group of hybrid lead halides, including those with 3D, 1D, and 0D inorganic substructures. Future advancements in this methodology will focus on predicting and decoding barcodes into potential sets of promising crystal structures.

Summing up, we show that the topological representation of crystal structures is suitable as a machine-readable representation of the organic-inorganic periodic structures for machine learning algorithms. Using this invariant representation, machine learning algorithms successfully predict the composition-structure-property relationships for LHP materials.

**Acknowledgments**

*This work was supported by the Interdisciplinary Scientific and Educational Schools of Lomonosov Moscow State University (grant № 23-Sh03-04).*

References:

1. J.C. Blancon, J. Even, C.C. Stoumpos, M.G. Kanatzidis, and A.D. Mohite. *Nat. Nanotechnol.,* 2020, 1512, **15**, pp. 969–985.

2. J. Huang, Y. Yuan, Y. Shao, and Y. Yan. *Nat. Rev. Mater.,* 2017, 27, **2**, pp. 1–19.

3. W. Li, Z. Wang, F. Deschler, S. Gao, R.H. Friend and A.K. Cheetham. *Nat. Rev. Mater.,* 2017, 23, **2**, pp. 1–18.

4. G. Grancini, M.K. Nazeeruddin. *Nat. Rev. Mater.*, 2019, **4**, pp. 4–22.

5. L. Mao, C.C. Stoumpos, M.G. Kanatzidis. *J. Am. Chem. Soc.*, 2019, **141**, pp.1171–1190.

6. E.I. Marchenko, V.V. Korolev, S.A. Fateev, A. Mitrofanov, N.N. Eremin, E.A. Goodilin, and A.B. Tarasov. *Chem. Mater.*, 2021, **33**, pp. 7518–7526.

7. Z. Wan, Q. De Wang, D. Liu, J. Liang. *New J. Chem.*, 2021, **45**, pp. 9427–9433.

8. E.I. Marchenko, S.A. Fateev, A.A. Petrov, V.V. Korolev, A. Mitrofanov, A.V. Petrov, E.A. Goodilin and A.B. Tarasov. *Chem. Mater.*, 2020, **32**, pp. 7383–7388.

9. W.B. Park, J. Chung, J. Jung, K. Sohn, S. P. Singh, M. Pyo, N. Shin, K.-S. Sohn. *IUCrJ*, 2017, **4**, pp. 486–494.

10. C.S. Hu, R. Mayengbam, M.C. Wu, K. Xia, T.C. Sum. *Commun. Mater.*, 2024, **5**, pp. 1–10.

11. E.I. Marchenko, V.V. korolev, A. Mitrofanov, S.A. Fateev, E.A. Goodilin, A.B. Tarasov. *Chem. Mater.*, 2021, **33**, pp. 1213–1217.

12. E.I. Marchenko, S.A. Fateev, A.A. Ordinartsev, P.A. Ivlev, E.A. Goodilin, A.B. Tarasov. *Mendeleev Commun.*, 2022, **32**, pp. 315–316.

13. Y. Jiang, D. Chen, X. Chen, T. Li, G.-W. Wei, F. Pan. *npj Comput. Mater.*, 2021, **7**, pp. 1–8.

14. B. Meredig, A. Agrawal, S. Kirklin, J.E. Saal, J.W. Doak, A. Thompson, K. Zhang, A. Choudhary, and C. Wolverton. *Phys. Rev. B - Condens. Matter Mater. Phys.*, 2014, **89**, pp. 1–7.

15. K. Momma, F. Izumi. *J. Appl. Crystallogr.*, 2011, **44**, pp. 1272–1276.

16. V.A. Blatov, A.P. Shevchenko, D.M. Proserpio. *Cryst. Growth Des.*, 2014, **14**, pp. 3576–3586.

17. E.I. Marchenko, S.A. fateev, V.V. Korolev, V. Buchinskiy, N.N. Eremin, E.A. Goodilin, and A.B. Tarasov. *J. Mater. Chem. C*, 2022, 44, 10, pp. 16838-16846.

18. S. Li, Y Liu, D. Chen, Y. Jiang, Z. Nie, F. Pan. *Wiley Interdiscip. Rev. Comput. Mol. Sci.*, 2022, **12**, pp. 1–20.

19. F. Pedregosa, G. Varoquaux, A. Gramfort, V. Michel, B. Thirion, O. Grisel, M. Blondel, P.

Prettenhofer, R. Weiss, V. Dubourg, J. Vanderplas, A. Passos, D. Cournapeau, M. Brucher, M. Perrot, É. Duchesnay. *J. Mach. Learn. Res.*, 2011, **12**, pp. 2825–2830.

20. L. Ward, A. Agrawal, A. Choudhary, C. Wolverton. *npj Comput. Mater.*, 2016, **2**, pp. 1–7.

## Graphical abstract

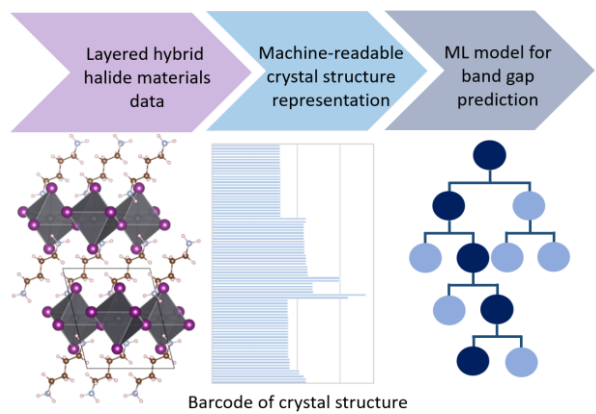